\documentclass{PoS}
\usepackage{epsf}
\usepackage{epsfig}

\newcommand{\coleps}{\epsilon_{I\alpha\beta}}
\newcommand{\flaeps}{\epsilon_{Iij}}
\newcommand{\setq}[2]{\{{\bf q}_{#1}^{#2}\}}

\newcommand{\intspace}[1]{\int d^4 {#1}\,}
\newcommand{\ha}{\frac{1}{2}}
\newcommand{\rr}{{\bf r}}
\newcommand{\q}[2]{ {\bf q}_{#1}^{#2}}
\newcommand{\vu}{{\bf u}}
\newcommand{\qia}{ {\bf q}_{I}^{a}}

\title{The rigidity of three flavor quark matter}

\ShortTitle{The rigidity of three flavor quark matter {\rm
\hspace*{3.8cm} Massimo Mannarelli and}}

\author{\speaker{Massimo Mannarelli}\\
        Instituto de Ciencias del Espacio (IEEC/CSIC) Campus Universitat Aut\`onoma de Barcelona, Facultat de Ci\`encies, Torre C5, E-08193 Bellaterra (Barcelona), Spain\\
        E-mail: \email{massimo@ieec.uab.es}}

\author{\speaker{Rishi Sharma}
         \thanks{Joint contribution of Massimo Mannarelli and Rishi Sharma. We thank Krishna Rajagopal
         for collaboration}\\
        Los Alamos National Laboratory\\
        E-mail: \email{rishi@lanl.gov}}

\abstract{Cold three flavor quark matter at large (but not
asymptotically large) densities may exist in a crystalline color
superconducting phase. These phases are characterized by a gap
parameter $\Delta$ that varies periodically in space, forming a
crystal structure. A Ginzburg-Landau expansion in $\Delta$ shows
that two crystal structures based on cubic symmetry are particularly
favorable, and may be the ground state of matter at densities
present in neutron star cores. We derive the effective action for
the phonon fields that describe space- and time-dependent
fluctuations of the crystal structure formed by $\Delta$, and obtain
the shear modulus from the coefficients of the spatial derivative
terms. Within a  Ginzburg-Landau approximation, we find shear moduli
which are 20 to 1000 times larger than those of neutron star crusts.
This phase of matter is thus more rigid than any known material in
the universe, but at the same time the crystalline color
superconducting phase is also superfluid. These properties raise the
possibility that the presence of this phase within neutron stars may
have distinct implications for their phenomenology. For example,
(some) pulsar glitches may originate in crystalline superconducting
neutron star cores. \vskip -510.truept \hspace*{10.cm} LA-UR-
08-08047 \vskip 510.truept}
\FullConference{8th Conference Quark Confinement and the Hadron Spectrum\\
                 September 1-6, 2008\\
                 Mainz. Germany}
\begin{document}

\section{Introduction}
At the extreme densities present at the cores of the neutron stars, quarks may be
deconfined. At present, the most promising direction to resolve whether
quark matter is present in these compact objects is to find
consequences of its presence for their phenomenology, which may allow
observations to rule out or in this possibility.

According with quantum-chromodynamics  (QCD), quark-quark
interaction is strong and attractive between quarks that are
antisymmetric in color, so we expect cold dense quark matter to be
in a color superconducting state. The critical temperatures of these
phases are generically on the order of tens of MeV, much larger than
the tens of keV temperatures of neutron star cores~\cite{reviews}.
This implies that if quark matter is present there, it must be in a
color superconducting phase, and also allows us to work at
temperature $T=0$ for the calculations we discuss below.

At asymptotic densities, where the up ($u$), down ($d$)
and strange ($s$) quarks can all be treated as massless, quark matter is in the
Color Flavor Locked (CFL) phase~\cite{Alford:1998mk,reviews}. The CFL condensate
is antisymmetric in color and flavor indices and therefore involves pairing
between quarks of different colors and flavors. All fermionic excitations are
gapped, with a gap parameter $\Delta_0\sim 10-100$~MeV.

However, at the cores of neutron stars, the quark number chemical
potential $\mu$ is expected to be between $350$ and $500$MeV which means that
$M_s$, lying somewhere between its current mass of order $100$~MeV and its
vacuum constituent mass of order $500$~MeV, cannot be neglected. In addition
bulk matter must be in weak equilibrium and must be electrically and color
neutral. These factors tend to separate the quark Fermi surfaces by $\sim
M_s^2/\mu$, and thus disfavor the cross-species BCS pairing which characterizes
the CFL phase. In neutral unpaired quark matter in weak equilibrium, to lowest
order in $M_s^2/\mu^2$, the quarks can be treated as if they were massless, but
with chemical potential splittings $\delta\mu_2\equiv (\mu_u-\mu_s)/2$ and
$\delta\mu_3\equiv (\mu_d -\mu_u)/2$ given by $\delta\mu_2=\delta\mu_3\equiv
\delta\mu = M_s^2/(8\mu)$.  Note that the splitting between unpaired Fermi
surfaces increases with decreasing density. In the CFL phase, the Fermi momenta
are {\it not} given by these optimal values for unpaired quark matter; instead,
the system pays a free energy price $\propto \delta\mu^2 \mu^2$ to equalize all
Fermi momenta and gains a pairing energy benefit $\propto \Delta_0^2\mu^2$.  As
a function of decreasing density, there comes a point (at which $\delta\mu
\approx \Delta_0/4$) when the system can lower its energy by breaking
pairs~\cite{Alford:2003fq}. Restricting the analysis to spatially homogeneous
condensates, the phase that results when CFL Cooper pairs start to break is the
gapless CFL  phase ~\cite{Alford:2003fq,Fukushima:2004zq}. However, this phase
turns out to be ``magnetically unstable''
\cite{Huang:2004bg,Casalbuoni:2004tb,Fukushima:2007bj}, meaning that it is
unstable to the formation of counter-propagating currents. If $\Delta_0$ is
small enough that there is a window of densities for which the gapless CFL phase
has a lower free energy than the CFL phase before nuclear matter takes over from
quark matter, then some other color superconducting phase(s) with a
smaller free energy than the gapless CFL phase must be the ground state of quark
matter in this window.

Assuming such a window exists, a possible resolution of the
``magnetic instability'', in particular for the lower values of
densities, is the crystalline color superconducting
phase~\cite{Alford:2000ze} which is the QCD analogue of a form of
non-BCS pairing first considered by Larkin, Ovchinnikov, Fulde and
Ferrell (LOFF)~\cite{LOFF}. (For two alternate possibilities
see~\cite{Kryjevski:2005qq} and~\cite{Gorbar:2005rx} .) This phase
is an attractive candidate in the intermediate density regime
because it allows pairing to occur even with split Fermi surfaces
in the free-energetically optimal way as in the absence of pairing,
by permitting Cooper pairs with non-zero net momentum. For example,
by allowing $u$ quarks with momentum ${\bf p+q_3}$ to pair with $d$
quarks with momentum ${\bf -p+q_3}$, for a fixed  ${\bf q_3}$, we
can pair $u$ and $d$ quarks along rings on their respective Fermi
surfaces~\cite{Alford:2000ze,Bowers:2002xr}. In coordinate space,
this corresponds to a condensate of the form $\langle ud \rangle\sim
\Delta_3 \exp\bigl({2i{\bf q_3}\cdot{\bf r}}\bigr)$.  The net free
energy gained due to pairing is then a balance between increasing
$|{\bf q_3}|$ yielding pairing on larger rings while exacting a
greater kinetic energy cost. The optimum choice turns out to be
$|{\bf q_3}|=\eta \delta\mu_3$ with $\eta=1.1997$, corresponding to
pairing rings on the Fermi surfaces with opening angle
$67.1^\circ$~\cite{Alford:2000ze}.  It is possible to cover larger
areas of the Fermi surfaces by allowing Cooper pairs with the same
$|{\bf q_3}|$ but various $\hat{\bf q}_3$, yielding $\langle ud
\rangle \sim \Delta_3 \sum_{\q{3}{a}}\exp\bigl(2\, i\, \q{3}{a}
\cdot {\bf r}\bigr)$ with the $\q{3}{a}$ chosen from some specified
set $\setq{3}{}$.  This is a condensate modulated in position space
in some crystalline pattern, with the crystal structure defined by
$\setq{3}{}$.  In this two-flavor context, a Ginzburg-Landau
analysis reveals that the best $\setq{3}{}$ contains eight vectors
pointing at the corners of a cube, say in the $(\pm 1,\pm 1,\pm 1)$
directions in momentum space, yielding a face-centered cubic
structure in position space~\cite{Bowers:2002xr}. In the following
section we generalize the pairing ansatz to the three flavor case.

\section{Three flavor crystalline color superconductivity}

In Refs.~\cite{Casalbuoni:2005zp,Mannarelli:2006fy,Rajagopal:2006ig}
the possibility that three flavor quark matter has a crystalline
color superconducting structure was explored. Considering the
pairing of the $u$ and $d$ quarks with the $s$ quark, we use a
pairing ansatz of form,
\begin{equation}
\langle\psi_{i\alpha}C\gamma^5\psi_{j\beta}\rangle\propto
\sum_I\coleps\flaeps\Delta_I\sum_{{\bf q}_I^a\in \setq{I}{}}\exp({2i{\bf q}_I^a\cdot{\bf r}})\ .
\label{condensate}
\end{equation}
This is antisymmetric in color ($\alpha,\beta$), spin,
and flavor ($i,j$) (where ($1$, $2$, $3$) correspond to ($u$, $d$, $s$)
respectively) and is thus a generalization
of the CFL condensate to crystalline color superconductivity.
For simplicity, we set $\Delta_1 = 0$, neglecting
$\langle ds \rangle$ pairing because the $d$ and $s$ Fermi
surfaces  are twice as far apart from each other as each is from
the intervening $u$ Fermi surface. Hence, the index $I$ can be taken to run over $2$ and $3$ only.
$\setq{2}{}$ ($\setq{3}{}$) defines the crystal structure of the $\langle us
\rangle$ ($\langle ud \rangle$) condensate.

We will analyze the system in an NJL model, which gives in the mean field
approximation an interaction term
\begin{equation}
{\cal L}_{\rm interaction}= \frac{1}{2} \bar{\psi}\Delta({\bf r})\bar{\psi}^T +
h.c.,
\label{meanfieldapprox}
\end{equation}
where the proportionality constant in~(\ref{condensate}) is conventionally
chosen so that
\begin{equation}
\Delta(\rr) = (C\gamma^5)
\sum_I\coleps\flaeps\Delta_I\sum_{{\bf q}_I^a\in\setq{I}{}}\exp({2i{\bf q}_I^a\cdot{\bf r}})
\label{precisecondensate}\;.
\end{equation}

The authors of~\cite{Rajagopal:2006ig} calculated the free energy $\Omega$ of
several crystalline structures within the weak coupling and Ginzburg-Landau
approximations, and found two qualitative rules that guide our understanding of
what crystal structures are favored in three flavor quark matter. First, the
$\langle ud \rangle$ condensates separately should be chosen to have favorable
free energies, as evaluated in the two flavor model of Ref.~\cite{Bowers:2002xr}.
Second, the $\langle ud \rangle$ and $\langle us \rangle$ condensates should be
rotated relative to each other in such a way as to maximize the angles between
the wave vectors describing the crystal structure of the $\langle ud \rangle$
condensate and the antipodes of the wave vectors describing the $\langle us
\rangle$ condensate.  This second qualitative rule can be understood as
minimizing the ``competition'' between the two condensates for up quarks on the
up Fermi surface~\cite{Rajagopal:2006ig}.

\begin{figure}[!ht]
\leavevmode
\begin{center}
\epsfysize=7cm
\epsfbox{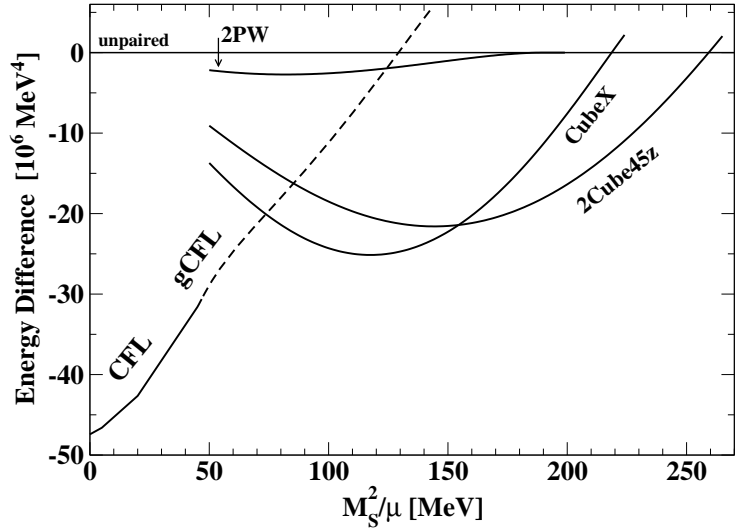}
\end{center}
\caption{The free energies of CubeX and 2Cube45z structures as a function of
$M_s^2/\mu$, which is proportional to the Fermi surface splitting. For comparison
we also show the free energy of the CFL phase, of the gapless CFL phase (gCFL) and of a crystalline condensate characterized by two antipodal plane waves (2PW).}
\label{omegabasic}
\end{figure}

 Two of the structures that possess these two features, called the CubeX
and 2Cube45z that we describe below, have a lower $\Omega$ than other considered structures. One or the other of the two, is also favored over unpaired quark
matter and the gapless CFL phase over a large parameter
range~\cite{Rajagopal:2006ig,Casalbuoni:2006zs}, see Fig.~\ref{omegabasic} .

Their robustness, however goes with a large gap parameter $\Delta$,
and the Ginzburg-Landau expansion parameter $(\Delta/\delta\mu)^2$
for these phases can be as large as a fourth for some values of
$\mu$, and hence the approximation is at the edge of its validity.
Nevertheless, their impressive robustness over a large range of
$\mu$ relevant for cores of neutron stars, make them well motivated
candidate phases to consider for phenomenological implications. For
analysis which go beyond the Ginzburg-Landau approximation
see~\cite{Mannarelli:2006fy,Nickel:2008ng}.

The CubeX crystal structure consists of eight vectors that belong to two sets of four vectors,
$\{{\bf \hat q_2}\}$  that can be taken as $\{({1}/{\sqrt{3}}) (\pm
\sqrt{2},0,\pm1)\}$, the four possible combinations of the signs giving the four
momentum directions, and $\{{\bf \hat q_3}\}$ that can be taken as
$\{({1}{\sqrt{3}}) (0,\pm\sqrt{2},\pm1)\}$.  The 2Cube45z crystal structure is
specified by two sets of eight unit vectors; the first set $\{{\bf \hat q_2}\}$  given by
 $\{(1/\sqrt{3})(\pm 1,\pm 1,\pm 1)\}$  and  the second one $\{{\bf \hat q_3}\}$ given by
$\{(1/\sqrt{3})(\pm \sqrt{2},0,\pm 1)\}\cup\{(1/\sqrt{3})(0,\pm \sqrt{2},\pm
1)\}$. For these structures, $\setq{2}{}$ can be transformed to $\setq{3}{}$ by
rigid rotations, ensuring that there are electrically neutral solutions of the
gap equation with $\Delta_2=\Delta_3=\Delta$~\cite{Rajagopal:2006ig}, a fact we
will use below.


\section{Phonons}

The crystalline phases of color superconducting quark matter that we have
described in the previous Section are unique among all forms of dense matter
that may arise within neutron star cores in one respect: they are
rigid~\cite{Mannarelli:2007bs}.  They are not solids in the usual sense: the
quarks are not fixed in place at the vertices of some crystal structure.
Instead, in fact, these phases are superfluid since the condensates all
spontaneously break the $U(1)_B$ symmetry corresponding to quark number.  The
diquark condensate, although spatially inhomogeneous, can carry
supercurrents~\cite{Alford:2000ze,Mannarelli:2007bs}.  And yet, we shall see
that crystalline color superconductors are rigid solids with large shear moduli.
It is the spatial modulation of the gap parameter that breaks translation
invariance, and it is this pattern of modulation that is rigid.  This novel form
of rigidity may sound tenuous upon first hearing, but we shall present the
effective Lagrangian that describes the phonons in the CubeX and 2Cube45z
crystalline phases, whose lowest order coefficients have  been calculated in the
NJL model that we are employing~\cite{Mannarelli:2007bs}.  We shall then extract
the shear moduli from the phonon effective action, quantifying the rigidity and
indicating the presence of transverse phonons.

Phonons in the crystal correspond to space- and time-varying displacements of the crystalline
pattern~\cite{Casalbuoni:2002my}.  In the present context, we introduce
displacement fields for the $\langle ud \rangle$, $\langle us \rangle$ and
$\langle ds \rangle$  condensates by making the replacement
\begin{equation}
\Delta_I \sum_{\q{I}{a}\in\setq{I}{}}e^{2i\q{I}{a}\cdot\rr} \rightarrow
\Delta_I \sum_{\q{I}{a}\in\setq{I}{}}e^{2i\q{I}{a}\cdot(\rr - \vu_I(\rr))}
\label{displacementfields}
\end{equation}
in (\ref{precisecondensate}).
One way to obtain the effective action describing the dynamics of the
displacement fields $\vu_I(\rr)$, including both its form and the values of its
coefficients within the NJL model that we are employing, is to take the mean
field NJL interaction to be given by~(\ref{meanfieldapprox}), but with
(\ref{displacementfields}), and integrate out the fermion fields. Since the
gapless fermions do not contribute to the shear modulus, one can integrate out
the fermions completely for this calculation. Note
that this is not true for the calculation of thermal or transport properties,
where the gapless fermions do contribute.

Upon carrying out the fermionic functional integration, we obtain,
\begin{eqnarray}
& &S[{\bf u}]=
\ha\intspace{x}\sum_I \kappa_I\label{Seff4}\\
& &\!\!\!\times\Biggl[
 \left(
   \sum_{\qia\in\setq{I}{}}(\hat{q}_I^a)^m(\hat{q}_I^a)^n \right)(\partial_0 u_I^m)(\partial_0 u_I^n)
-\left(
   \sum_{\qia\in\setq{I}{}}(\hat{q}_I^a)^m(\hat{q}_I^a)^v(\hat{q}_I^a)^n(\hat{q}_I^a)^w\right)
   (\partial_v u_I^m)(\partial_w u_I^n)
\Biggr] \nonumber\;
\end{eqnarray}
where $m$, $n$, $v$ and $w$ are spatial indices running over $x$, $y$ and $z$
and where we have defined
\begin{equation}
\kappa_I\equiv
\frac{2\mu^2|\Delta_I|^2\eta^2}{\pi^2(\eta^2-1)} \;.
\label{lambdapw}
\end{equation}
For $\Delta_1=0,\;\;\Delta_2=\Delta_3=\Delta,{\mbox{ and }}\eta\simeq1.1997,$
\begin{equation}
\kappa_2=\kappa_3\equiv\kappa\simeq 0.664\,\mu^2|\Delta^2|\;.\label{kappavalue}
\end{equation}

$S[{\bf u}]$  is the low energy effective action for phonons in any
crystalline color superconducting phase, valid to second order in
derivatives, to second order in the gap parameters $\Delta_I$ and to
second order in the phonon fields $\vu_I$.    Because we are
interested in long wavelength, small amplitude, phonon excitations,
expanding to second order in derivatives and in the phonon fields is
satisfactory. The Ginzburg-Landau expansion, which gives a series in
$(\Delta/\delta\mu)^2$, is not under quantitative control for these
most favorable phases, as we discussed in the previous Section. But
as we shall see, for glitch phenomenology, the main requirement from
the shear modulus calculation is that it should be large, and given
that we get much larger values than those obtained for conventional
neutron star crusts, there is at present no great motivation to go
to higher orders. At this order in $(\Delta_I)^2$, there is no
mixing between different $\vu_I$, and they can all be treated
independently.

In order to extract the shear moduli, we need to compare the phonon
effective action to the 
theory of elastic media~\cite{Landau:Elastic},
which requires introducing the strain tensor
\begin{equation}
s_I^{mv}\equiv\ha\Bigl(\frac{\partial  u_I^m}{\partial
x^v}+\frac{\partial  u_I^v}{\partial x^m}\Bigr).\label{strain}
\end{equation}
We then wish to compare the action (\ref{Seff4}) to
\begin{equation}
{S}[{\bf u}]= \ha\intspace{x}\Biggl(
    \sum_I\sum_m \rho_I^m (\partial_0  u_I^m)(\partial_0  u_I^m)
    -\sum_{I}\sum_{{mn}\atop{vw}}\lambda_I^{mvnw}
    s_I^{mv}s_I^{nw}\Biggr)
\label{full action},
\end{equation}
which is the general form of the action for $\vu_I$ that don't mix, in the case
in which the effective action is quadratic in displacements and which defines
the elastic modulus tensor $\lambda_I^{mvnw}$ for this case.  In this case, the
stress tensor (in general the derivative of the potential energy with respect to
$s_I^{mv}$) is given by
\begin{equation}
\sigma_I^{mv}
=
\lambda_I^{mvnw}s_I^{nw}\label{stress2}\; .
\end{equation}
The diagonal components of $\sigma$ are proportional to the compression exerted
on the system and are therefore related to  the bulk  modulus of the crystalline
color superconducting quark matter. Since unpaired quark matter  has a pressure
$\sim \mu^4$, it gives a contribution to the bulk modulus that  completely
overwhelms the contribution from the condensation into a crystalline phase,
which is of order $\mu^2\Delta^2$.  We shall therefore not calculate the bulk
modulus.  On the other hand, the response to shear stress arises only because of
the presence  of the crystalline condensate.  The shear modulus is defined as
follows. Imagine exerting a static external stress $\sigma_I$ having only an
off-diagonal component, meaning  $\sigma^{mv}_I\neq 0$ for a pair of space
directions $m\neq v$, and all the other components of $\sigma$ are zero. The
system will respond with a strain $s_I^{nw}$.  The shear modulus in the $mv$
plane is then
\begin{equation}
\nu_I^{mv} \equiv \frac{\sigma_I^{mv}}{2s_I^{mv}}
= \ha\lambda_I^{mvmv}
\label{defineshearmodulus}\;,
\end{equation}
where the indices $m$ and $v$ are not summed.
For a general quadratic potential with $\sigma_I^{mv}$ given by (\ref{stress2}),
$\nu_I^{mv}$ simplifies partially but the full simplification given by the last
equality in  (\ref{defineshearmodulus}) only arises for special cases in which
the only nonzero entries in $\lambda^{mvnw}$ with $m\neq v$ are the
$\lambda^{mvmv}$ entries, as is the case for all the crystal structures that we
consider.

For a given crystal structure, upon evaluating the sums in (\ref{Seff4}) and
then using the definition (\ref{strain}) to compare (\ref{Seff4}) to (\ref{full
action}), we can extract expressions for the $\lambda$ tensor and thence for the
shear moduli.  This analysis, described in detail in \cite{Mannarelli:2007bs},
shows that in the CubeX phase
\begin{equation}
\nu_2=\frac{16}{9}\kappa\left( \begin{array}{ccc}
0 & 0 & 1\\
0 & 0 & 0\\
1 & 0 & 0
\end{array}
\right)\,,\hspace{.3cm}  \nu_3=\frac{16}{9}\kappa\left(
\begin{array}{ccc}
0 & 0 & 0\\
0 & 0 & 1\\
0 & 1 & 0
\end{array}
\right)\label{nu2 and nu3}\;,
\end{equation}
while in the 2Cube45z phase
\begin{equation}
\nu_{2}=\frac{16}{9}\kappa\left( \begin{array}{ccc}
0 & 1 & 1\\
1 & 0 & 1\\
1 & 1 & 0
\end{array}
\right)\,,\ \
\nu_{3}=\frac{16}{9}\kappa\left( \begin{array}{ccc}
0 & 0 & 1\\
0 & 0 & 1\\
1 & 1 & 0
\end{array}
\right)\label{nu 2Cube45z}\;.
\end{equation}
We shall see in the next Section that it is relevant for glitch phenomenology to check that both these
crystals have enough nonzero entries in their shear moduli $\nu_I$ that if there
are rotational vortices are pinned within them, a force seeking to move such a
vortex is opposed by the rigidity of the crystal structure described by one or
more of the nonzero entries in the $\nu_I$.  This is demonstrated in
\cite{Mannarelli:2007bs}.

We see that all the nonzero shear moduli of both the CubeX and 2Cube45z
crystalline color superconducting phases turn out to take on the same value,
\begin{equation}
\nu_{\rm CQM} = \frac{16}{9}\kappa
=  1.18\, \mu^2 \Delta^2
=  2.47\, \frac{{\rm MeV}}{{\rm fm}^3}
\left(\frac{\Delta}{10~{\rm MeV}}\right)^2 \left(\frac{\mu}{400~\rm{MeV}}\right)^2,
\label{shearmodulus}
\end{equation}
where $\mu$ is expected to lie between $350$ to $500$MeV and $\Delta$ may be
taken to be between $5$ and $25$MeV to obtain numerical estimates.

From (\ref{shearmodulus}) we first of all see that the shear modulus is in no
way suppressed relative to the scale  $\mu^2\Delta^2$ that could have been
guessed on dimensional grounds.  And, second, we discover that a quark matter
core in a crystalline color superconducting phase is 20 to 1000 times more
rigid than the crust of a conventional neutron star~\cite{Strohmayer:1991}.
Finally,  one can extract the phonon dispersion relations from the effective
action~(\ref{Seff4}).  The transverse phonons, whose restoring force is
provided by the shear modulus turn out to have direction-dependent velocities
that are typically a substantial fraction of the speed of light, in the
specific instances evaluated in \cite{Mannarelli:2007bs} being given by
$\sqrt{1/3}$ and $\sqrt{2/3}$.  This is yet a third way of seeing that this
superfluid phase of matter is rigid indeed.

\section{Rigid quark matter}

The existence of a rigid crystalline color superconducting core within neutron
stars may have a variety of observable consequences.   For example, if some
agency ({\it e.g.} magnetic fields not aligned with the rotation axis) could
maintain the rigid core in a shape that has a nonzero quadrupole moment, gravity
waves would be emitted.  The LIGO non-detection of such gravity waves from
nearby neutron stars already limits the possibility that they have rigid cores
that are deformed to the maximum extent allowed by the shear modulus
(\ref{shearmodulus})~\cite{Abbott:2007ce,Haskell:2007sh,Lin:2007rz}.  Perhaps the most
exciting implication of a rigid core, however, is the possibility that (some)
pulsar ``glitches'' could originate deep within a neutron star, in its quark
matter core.

A spinning neutron star observed as a  pulsar gradually spins down as it loses
rotational energy to electromagnetic radiation.  But, every once in a while the
angular velocity at the crust of the star is observed to increase suddenly in a
dramatic event called a glitch.  The standard explanation~\cite{Anderson:1975}
(see~\cite{Mannarelli:2007bs} for more Refs.) requires the presence of a
superfluid in some region of the star which also features a rigid structure that
can pin the vortices in the rotating superfluid and that does not easily deform
when the vortices pinned to it are under tension.

As a spinning pulsar slowly loses angular momentum over years, since the angular
momentum of any superfluid component of the star is proportional to the density
of vortices, the vortices ``want'' to move apart.  However, if  the vortices are
pinned to a rigid structure, these vortices do not move and after a time this
superfluid component of the star is spinning faster than the rest of the star.
When the ``tension'' built up in the array of pinned vortices reaches a critical
value, there is a sudden ``avalanche'' in which vortices unpin, move outwards
reducing the angular momentum of the superfluid, {\it and then re-pin}.  As this
superfluid suddenly loses angular momentum, the rest of the star, including in
particular the surface whose angular velocity is observed, speeds up --- a
glitch.
In the standard explanation of pulsar glitches, this occurs in the
inner crust of a neutron star where a neutron superfluid coexists
with a rigid array of positively charged nuclei that may serve as vortex pinning
sites.  In recent work, Link has concluded that this scenario is not viable
because once neutron vortices are moving through the inner crust, as must happen
during a glitch, they  are so resistant to bending that they can never
re-pin~\cite{LinkPrivateCommunication}.  Link concludes that we do not currently
understand the origin of glitches as a crustal phenomenon.

By virtue of being simultaneously superfluids and rigid solids, the crystalline
phases of quark matter provide all the necessary conditions to be the locus in
which (some) pulsar glitches originate.
Their shear moduli
(\ref{shearmodulus}),
makes them more than rigid enough for glitches to originate within them.  The
crystalline phases are at the same time superfluid, and it is reasonable to
expect that the superfluid vortices  will have lower
free energy if they are centered along the intersections of the nodal planes of
the underlying crystal structure, {\it i.e.}  along lines along which the condensate
already vanishes even in the absence of a rotational vortex. A crude estimate of
the pinning force on vortices within crystalline color superconducting quark
matter indicates that it is sufficient~\cite{Mannarelli:2007bs}.  So, the basic
requirements for superfluid vortices pinning to a rigid structure are all
present.  The central questions that remain to be addressed are the explicit
construction of vortices in the crystalline phase and the calculation of their
pinning force, as well as the calculation of the timescale over which sudden
changes in the angular momentum of the core are communicated to the (observed)
surface, presumably either via the common electron fluid or via magnetic
stresses.

Much theoretical work remains before the hypothesis that pulsar glitches
originate within a crystalline color superconducting neutron star core is
developed fully enough to allow it to confront data on the magnitudes,
relaxation timescales, and repeat rates that characterize the data.
Nevertheless, this hypothesis offers one immediate advantage over the
conventional scenario that relied on vortex pinning in the neutron star crust.
Link has observed that it is impossible for a neutron star anywhere within which
rotational vortices are pinned to precess~\cite{Link:2006nc}, and yet there is
now evidence that several pulsars are
precessing~\cite{Stairs:2000}. 
Since {\it all} neutron stars have crusts, the precession of any pulsar is inconsistent with the
pinning of vortices within the crust, a requirement in the standard explanation
of glitches.  On the other hand, perhaps not all neutron stars have crystalline
quark matter cores --- for example, perhaps the lightest neutron stars have
nuclear matter cores.  Then, if vortices are never pinned in the crust but are
pinned within a crystalline quark matter core, those neutron stars that do have
a crystalline quark matter core can glitch but cannot precess while those that
don't can precess but cannot glitch.

\section*{Acknowledgements}
Massimo Mannarelli and Rishi Sharma would like to thank Krishna
Rajagopal for collaboration on the work presented here. They would
also like to thank the organizers of the conference ``Confinement
2008'' for giving us an opportunity to present our work. MM has been
supported by the ``Bruno Rossi" fellowship program and by the
Spanish grant AYA 2005-08013-C03-02.   This research was supported
in part by the Office of Nuclear Physics of the Office of Science of
the U.S.~DoE, contract \#DE-AC02-05CH11231, cooperative research
agreement \#DF-FC02-94ER40818 and LANS, LLC for the NNSA of the DoE,
contract \#DE-AC52-06NA25396.

\end{document}